\documentstyle[12pt,epsfig]{article}
\def\bbuildrel#1_#2^#3%
{\mathrel{\mathop{\kern 0pt#1}\limits_{#2}^{#3}}}
\newcommand{\beq}{\begin{equation}}
\newcommand{\eeq}{\end{equation}}
\newcommand{\api}{\frac{\alpha_s}{\pi}}
\newcommand{\ba}{\begin{array}}
\newcommand{\ea}{\end{array}}
\newcommand{\as}{\alpha_s}

\newcommand{\msbar}{\overline{\mbox{MS}}}

\newcommand{\EQN}{\label}

\begin{document}

\begin{titlepage}
\noindent
%
%
\hfill TTP94--14\\
\mbox{}
\hfill September 1994   \\   
%
%
\vspace{0.5cm}
\begin{center}
  \begin{Large}
  \begin{bf}
${\cal O}(\alpha\as \ln m_t^2)$ Non-Universal Corrections
       to the Decay Rate
 $\Gamma(Z\rightarrow b\bar{b})$\footnote[1]
{Talks presented by:
   \begin{minipage}[t]{8cm}
        M. Steinhauser, Fr\"uhjahrstagung der DPG,
          1-4 March 1994, Dortmund; \\
        A. Kwiatkowski, CERN Workshop, 13 June 1994, CERN,
            Geneva.
   \end{minipage}     }
  \\
  \end{bf}
  \end{Large}
%
%

  \begin{large}
 A.Kwiatkowski\footnote[2]{Address after 15. October:
   \begin{minipage}[t]{8cm}
              Lawrence Berkeley Laboratory \\
              Physics Division \\
              Theoretical Physics Group \\
              1 Cyclotron Road \\
              Berkeley, CA 94720, USA
   \end{minipage}     }
                               , M.Steinhauser    \\[3mm]
    Institut f\"ur Theoretische Teilchenphysik\\
    Universit\"at Karlsruhe\\
    Kaiserstr. 12,    Postfach 6980\\[2mm]
    D-76128 Karlsruhe, Germany\\   \end{large}

%
%
  \vspace{1.5cm}
  {\bf Abstract}
\end{center}
\begin{quotation}
\noindent
The partial decay rate $\Gamma(Z\to b\bar{b})$ is significantly
influenced by the mass of the top quark due to electroweak radiative
corrections. The leading $\sim m_t^2$ and the next-to-leading
contribution $\sim \ln m_t^2$ are known to be numerically
of similar size. In this work we calculate the QCD corrections to the
logarithmic correction using the heavy top mass expansion.
The ${\cal O} (\alpha\as \ln m_t^2)$ corrections are of the same order
as the QCD corrections to the quadratic top mass term, but of different sign.
\end{quotation}
\end{titlepage}

\renewcommand{\arraystretch}{2}

\section{Introduction}

Although
the direct observation of the top quark is out of range for the experiments at
LEP, several observables are affected by the top quark through virtual
states in higher order radiative corrections.

High precision measurements and the comparison of these quantities
with the theoretical predictions allow to extract bounds on the
top mass. Present analysis of $e^+ e^-$ collisions at the $Z$ peak
estimates the top mass in the range $m_t=173^{+12 +13}_{+18 -21}$ GeV
\cite{Sch94}, which is in agreement with
top masses of $m_t=174\pm10^{+13}_{-12}$ GeV from
$\bar{p}p$ collisions at TEVATRON \cite{CDF}.
Of particular interest for deducing the limits on $m_t$ from LEP data
is the partial decay rate $\Gamma(Z\to b\bar{b})$.
On the one hand this quantity
exhibits a strong sensitivity on the top mass
\cite{AkhBarRie86}
as the leading term of the electroweak corrections is quadratic in $m_t$
and the next-to-leading logarithmic contribution $\ln m_t^2$
is numerically of the same size. On the other hand the already small
uncertainty of the measurement of
$R_{b\bar{b}}=\Gamma(Z\to b\bar{b})/\Gamma(Z\to \mbox{hadrons})
            =0.2192\pm0.0018$
\cite{Sch94} is expected to be reduced below 1\% in the future.
As a consequence the determination of QCD corrections to
the electroweak one loop result became increasingly important.

For the quadratic top mass contribution these
${\cal O} (\alpha\alpha_s m_t^2)$ corrections were calculated
by four independent groups
\cite{FleJegRacTar92,BucBur92,Deg93,CheKwiSte93}.
In this work we present the QCD corrections of order
${\cal O} (\alpha\alpha_s \ln m_t^2)$ to the next-to-leading
top mass contribution.

Our calculation is performed using dimensional regularisation in the
$\msbar$-scheme with anticommuting $\gamma_5$ in the 't Hooft-Feynman gauge.
As in our previous
work \cite{CheKwiSte93} we employ the hard mass procedure
\cite{PivTka84,GorLar87,CheSmi87,Smi91} to derive the
next-to-leading order in the inverse top mass expansion. In order to
handle large expressions during this expansion we use the
algebraic manipulation program FORM \cite{Verm}.
Massless multiloop integrals are evaluated with the help of the software
package MINCER \cite{LarTkaVer91}.

If we consider the properly resummed propagator of the $Z$ boson we
get in the case $q^2 \approx M_Z^2$ for the transverse part
\beq
D^{\mu\nu}_T(q^2) =
-\frac{1}{i}\frac{1}{1+\Pi^{\prime}(M_Z^2)}
 \frac{g_{\mu\nu}}{M_Z^2-q^2-\frac{i\mbox{\footnotesize Im }\Pi(M_Z^2)M_Z}
                                  {1+\Pi^{\prime}(M_Z^2)}}
\eeq
where $M_Z$ is the renormalized $Z$ boson mass and
$\Pi(q^2)$ is the transverse part of the polarisation tensor
$\Pi_{\mu\nu}(q^2)$. One can get $\Pi(q^2)$ through contracting with
$1/(D-1)(g_{\mu\nu}-q_{\mu}q_{\nu}/q^2)$ where $D=4-2\epsilon$
is the space-time dimension.
$\Pi^{\prime}(M_Z^2)$ is defined as
$d\mbox{ Re}\Pi(q^2)/dq^2 $ also evaluated for
$q^2=M_Z^2$.
Thus we can write the partial decay rate
$\Gamma(Z\rightarrow b\bar{b})$
in the following form
\beq
\Gamma(Z\rightarrow b\bar{b})=
      \bigg[ \Gamma_0\left(v^2+a^2\right) + \Delta\Gamma \bigg]
                              \frac{1}{1+\Pi^{\prime}(M_Z^2)}
\eeq
with $v=-1+\frac{4}{3}s_W^2, a=-1$ and $\Gamma_0=\alpha M_Z/48 c_W^2 s_W^2$.
$\Delta\Gamma$ contains all vertex corrections and the $\gamma Z$-mixing.
Expressing the result in terms of $G_F$ (instead of $\alpha$) introduces a
factor $(1-\Delta r)$.
The above equation can be written in the form
\beq
\Gamma(Z\rightarrow b\bar{b})=
        \bigg[ \frac{G_F M_Z^3}{8\sqrt{2}\pi}\left(v^2+a^2\right)
              +\Delta\Gamma \bigg]
                         \frac{1-\Delta r}{1+\Pi^{\prime}(M_Z^2)}.
\eeq
The top mass dependence of the last term is of universal nature.
This contribution --- as well as the universal part of $\Delta\Gamma$ ---
can be expressed through the universal $\rho$-parameter
\cite{DjoVer87Djo88,HalKniSto92}.
In this work we calculate the non-universal part of order
${\cal O}(\alpha\alpha_s\ln m_t^2)$.

The outline of the paper is as follows: Section 2 contains further details of
the calculation and in section 3 we present the results in the $\msbar$ and
OS-scheme.

\section{Calculation of the Order ${\cal O}(\alpha\as \ln m_t^2)$}

In the decay rate $\Gamma(Z\rightarrow b\bar{b})$ the top mass
dependence is due to the appearance of the top quark as a virtual
particle, i.e.
it is induced by $t-b$-transitions due to the exchange of
charged Higgs ghosts $\Phi^{\pm}$ or $W$ bosons.
One can distinguish between seven
classes of diagrams as listed in figure
\ref{figclasses} from which the imaginary part has to be calculated.
QCD corrections are obtained by attaching gluon lines
in all possible ways. This results in 58 topologically different diagrams.

\begin{figure}[b]
 \begin{center}
  \begin{tabular}{ccc}
  \\
  \\
  &
  \end{tabular}
 \caption{\label{figclasses}
          The seven distinct classes of electroweak diagrams, which
          contain the top quark.
          Dashed lines: Higgs ghost; wavy (internal) lines: $W$ boson;
          thin lines: $b$ quark; thick lines: $t$ quark.}
 \end{center}
\end{figure}

We perform our calculation in the large top mass limit $m_t\to \infty$,
i.e. we apply an asymptotic expansion in the inverse heavy mass
$1/m_t$ \cite{PivTka84,GorLar87,CheSmi87,Smi91}.
In practice one isolates all
possible hard subgraphs and expands them w.r.t. the
small masses and (external) momenta. Afterwards this expansion is inserted
as an effective vertex into the remaining diagram.

The hard mass procedure was already used in our previous work
\cite{CheKwiSte93} to evaluate the ${\cal O}(\alpha\alpha_s m_t^2)$
corrections. In this order only diagrams with Higgs ghost exchange
contributed.
The calculation of the next-to-leading
${\cal O}(\alpha\alpha_s \ln m_t^2)$
corrections is extended in three different ways. First,
the diagrams with $W$ boson exchange need to be taken into account.
Second, additional hard subgraphs of the Higgs ghost diagrams contribute
to the considered order. Third, the subgraphs already considered
in \cite{CheKwiSte93} must be expanded to the next higher order.

We find that the next-to-leading order results are finite separately
for each class of diagrams which contain a Higgs ghost. Third order
poles $1/\epsilon^3$ compensate in the sum of all possible hard subgraphs
of a diagram and second order poles $1/\epsilon^2$ cancel after
replacing the bare top mass by the $\msbar$ renormalized one
\beq
m_t^{\mbox{\tiny bare}}
      =\bar{m}_t \left( 1-\frac{\alpha_s}{\pi} \frac{1}{\epsilon} \right).
\eeq
The remaining simple divergence drops out in the imaginary part of
the diagram.

The class of $W$ diagrams and their renormalized contribution
$R\Pi^{{\tiny (W)}}(q^2)$
remain to be calculated. Here $R$ denotes the so-called Bogoliubov-Parasiuk
$R$-operation \cite{BogPar57}
adopted to the minimal subtraction scheme \cite{Col75}.
This procedure determines the finite renormalized value of a given
regularized Feynman integral. It works on a graph-by-graph
basis and removes all subdivergencies together with the overall
UV divergence in a way compatible with adding local counterterms to
the Lagrangian.

We apply this prescription to subtract all subdivergencies
of each $W$ diagram, whereas the overall divergence again is eliminated by
taking the imaginary part.

The pole part $\Delta(\gamma)$ of a divergent subgraph $\gamma$ of a
Feynman diagram $\Gamma$ is defined as the overall divergence of the subgraph
$\gamma$ and can be written as a polynomial in possible masses and external
momenta of $\gamma$ with coefficients being pure poles in $\epsilon$.
Since the $\epsilon$ poles themselves are independent of masses and momenta,
their determination can be simplified by setting masses or momenta zero.
However, care must be taken to guaranty that no spurious IR
singularities are introduced this way.
The subtracted Feynman integral
 $\Delta(\gamma)\langle\Gamma/\gamma\rangle$ is
obtained by replacing the subgraph $\gamma$ through $\Delta(\gamma)$.
For the calculation of the remaining integrals
 $\langle\Gamma/\gamma\rangle$
the heavy mass expansion can be employed.

There are several checks for the correctness of our result.
The first one is the fact that after renomalization the seven classes of
diagrams are separately finite.

The second check is the gauge invariance with respect to QCD.
The calculation is done with an arbitrary QCD gauge parameter $\xi_S$ which is
introduced through the gluon propagator. As expected each class is separately
independent of $\xi_S$ if the sum of all diagrams is taken.

The third check is the invariance of the gauge parameter
$\xi_W$ from the elektroweak theory. $\xi_W$ appears through the propagator
of the $W$ boson and the $\Phi^\pm$.
The total sum of all diagrams is indeed independent of
$\xi_W$.

\section{Results and Discussion}

In order to calculate explicitely the ${\cal O}(\alpha\alpha_s\ln m_t^2)$
correction to the $Z\to b\bar{b}$ vertex, we have employed the hard mass
procedure as an expansion up to the next-to-leading order in the
$1/m_t$ series.
As a consequence the leading order result is automatically reproduced.
Furthermore it became necessary to repeat the calculation up to
next-to-leading order for the case without QCD corrections
(see \cite{diplms}) since they induce corrections in first order
$\alpha_s$ through top mass renormalization.

Let us first recall this purely electroweak result for the
non-universal decay rate.

\begin{eqnarray}\EQN{num1}
\Delta\Gamma^{{\rm \tiny non-univ.}}_{\tiny Z\to b\bar{b}} & = &
-\Gamma_0\frac{G_F}{\sqrt{2}\pi^2}
\left( 1-\frac{2}{3}s_W^2 \right)
\bigg\{
      \bar{m}_t^2                                         \nonumber \\
& &
    + M_Z^2 \bigg[\ln\frac{\mu^2}{\bar{m}_t^2}
                \left(-3 + 6 s_W^2 - 3 s_W^4 \right)
                                                          \nonumber \\
& &
\hphantom{+M_Z^2}
                +\ln\frac{M_W^2}{\bar{m}_t^2}
                  \left(\frac{1}{6}-\frac{10}{3}s_W^2+3s_W^4\right)  \bigg]\\
&\stackrel{\mu^2=M_W^2}{=}&
-\Gamma_0\frac{G_F}{\sqrt{2}\pi^2}
\left( 1-\frac{2}{3}s_W^2 \right)                          \nonumber\\
& &
\hphantom{-\Gamma_0\frac{G_F}{\sqrt{2}\pi}}
\bigg\{
      \bar{m}_t^2
    + M_Z^2 \ln\frac{M_W^2}{\bar{m}_t^2}
                 \left(-\frac{17}{6}+\frac{8}{3}s_W^2 \right)
\bigg\}
\end{eqnarray}

The logarithms have their origins in different classes of diagrams.
{}From the diagrams with Higgs ghost exchange logarithms
$\ln\mu^2/\bar{m}_t^2$ and $\ln\mu^2/M_W^2$ occur such that in the
combination the $\mu$ drops out and $\ln M_W^2/\bar{m}_t^2$ are left.
The $\ln \mu^2/\bar{m}_t^2$ only comes from the $W$ graphs.
Choosing the scale $\mu^2=M_W^2$ reproduces the result known from
\cite{AkhBarRie86}.
Since we are only interested in top mass effects, i.e. in the difference
of a heavy and a light top quark, we are allowed to set $\mu^2=M_W^2$
because the difference
$
 \Gamma(Z\to b\bar{b})_{\mbox{\tiny heavy top}}
-\Gamma(Z\to b\bar{b})_{\mbox{\tiny light top}}
$
is independent of $\mu$ and only contains logarithms of the form
$\ln m_t^2/M_W^2$.

Including first order QCD corrections we obtain
the non-universal part of $\Delta\Gamma$ in the $\msbar$ scheme:

\begin{eqnarray}
\Delta\Gamma^{{\rm \tiny non-univ.}}_{\tiny Z\to b\bar{b}}  & = &
-\Gamma_0\frac{G_F}{\sqrt{2}\pi^2}
\left( 1-\frac{2}{3}s_W^2 \right)
   \frac{\as(\mu^2)}{\pi}   \nonumber \\
& &
M_Z^2\bigg\{
                \ln\frac{\mu^2}{\bar{m}_t^2}
                    \left( \frac{8}{3}+\frac{2}{3}s_W^2-3s_W^4 \right)
                +\frac{7}{81} \ln\frac{M_Z^2}{\bar{m}_t^2}
                    \left(-1+\frac{2}{3}s_W^2 \right)            \nonumber \\
& &
\hphantom{+M_Z^2}
               +\ln\frac{M_W^2}{\bar{m}_t^2}
                    \left( \frac{1}{6}-\frac{10}{3}s_W^2+3s_W^4 \right)\\
& &
\hphantom{+M_Z^2}
               +\Delta\Gamma_{\mbox{\tiny rem}}^{{\tiny\msbar}}  \nonumber
\bigg\}.
\end{eqnarray}

$\Delta\Gamma_{\mbox{\tiny rem}}$ contains the constant terms of the
non-universal diagrams containing the top quark and also terms from graphs
without top quark which we have not calculated.
Via the relation
\beq\EQN{r3}
\bar{m}(\mu^2) = m_{OS}\left(1-\api\left[\frac{4}{3}+\ln\frac{\mu^2}{m_{OS}^2}
\right]\right)
\eeq
we transform this result into the OS-scheme:

\begin{eqnarray}\EQN{num2}
\Delta\Gamma^{{\rm \tiny non-univ.}}_{\tiny Z\to b\bar{b}}  & = &
-\Gamma_0\frac{G_F}{\sqrt{2}\pi^2}
\left( 1-\frac{2}{3}s_W^2 \right)
    \frac{\as(\mu^2)}{\pi}   \nonumber \\
& &
M_Z^2 \bigg\{
                 \ln\frac{\mu^2}{m_t^2}
                        \left( -3 + 6 s_W^2-3s_W^4 \right) \nonumber \\
& &
\hphantom{M_Z^2}
                 +\frac{7}{81} \ln\frac{M_Z^2}{m_t^2}
                        \left(-1+\frac{2}{3}s_W^2 \right)\\
& &
\hphantom{M_Z^2}
                +\ln\frac{M_W^2}{m_t^2}
                       \left( \frac{1}{6}-\frac{10}{3}s_W^2+3s_W^4 \right)
                +\Delta\Gamma_{\mbox{\tiny rem}}^{{\tiny OS}}
\bigg\}                                   \nonumber\\
&\stackrel{\mu^2=M_W^2}{=}&
-\Gamma_0\frac{G_F}{\sqrt{2}\pi^2}
\left( 1-\frac{2}{3}s_W^2 \right)
    \frac{\as}{\pi}          \nonumber \\
& &
M_Z^2 \bigg\{
                 \ln\frac{M_W^2}{m_t^2}
                      \left( -\frac{17}{6}+\frac{8}{3}s_W^2 \right)  \\
& &
\hphantom{M_Z^2}
                 +\frac{7}{81} \ln\frac{M_Z^2}{m_t^2}
                      \left(-1+\frac{2}{3}s_W^2 \right)
                 +\Delta\Gamma_{\mbox{\tiny rem}}^{{\tiny OS}} \nonumber
\bigg\}
\end{eqnarray}

As a consequence
of dimensional regularisation the renormalization scale $\mu$ appears
in the calculation. For the diagrams with Higgs ghost exchange
again the logarithms $\ln \mu^2/m_t^2,
\ln \mu^2/M_W^2$ and $\ln \mu^2/M_Z^2$ are combined such that $\mu^2$
disappears. Since these
logarithms are connected with the pole structure of the result, this
cancellation is expected from the finiteness of these classes of diagrams.

For the classes of the $W$ boson exchange graphs with their explicit
calculation of counterterms in the $\msbar$ scheme a logarithm
$\ln \mu^2/\bar{m}_t^2$ remains left.

One can observe that
for the QCD corrections the
  coefficients of the $\ln \mu^2/\bar{m}_t^2$
and  the $\ln M_W^2/\bar{m}_t^2$ terms
in eq.(\ref{num2}) are identically the
same as in the pure electroweak result
of eq.(\ref{num1}).
Thus these logarithms are characterized by the
same correction factor $(1+\as/\pi)$ as it is known from the
pure QCD corrections to the Born decay rate.
With the electroweak result
being well established in the literature, the explicit
$\mu$ dependence in the OS result
 eq.(\ref{num1}) is known to be
compensated by the running electroweak running
coupling constant, thus giving a
RG invariant result. The common correction factor
$(1+\as/\pi)$ therefore also guaranties the
$\mu$-cancellation in our ${\cal O}(\alpha\as\ln m_t^2)$ result.

Having gained more experience and insight in the heavy mass expansion
as a practical approach of an effective theory since our previous
work \cite{CheKwiSte93}, let us add an additional comment concerning
the appropriate choice of scale as argument for the running mass and coupling
constant.

The leading ${\cal O}(\alpha\alpha_s m_t^2)$ calculation
exhibits a
remarkable structure.
All integrals factorize and can be classified into two cathegories, both
of which are separately finite and gauge invariant. The first one is
characterized by a factorization into a two loop massive tadpole
integral and a massless one loop p-integral, where the gluon is part
of the former. Since this integration is affected by only one mass
scale namely the top quark mass, it is
instructive
to evaluate $\alpha_s$ at the scale $m_t^2$ for these contributions.
Contrary to this the second cathegory comprises all corrections which
factorize into a one loop massive tadpole and a two loop massless p-integral,
with the gluon lines contained in the latter.
The only scale in the massless p-integral is the energy regime of the
process under consideration and it seems to be appropriate to
calculate $\alpha_s$ at this scale.

The result for the ${\cal O}(\alpha\alpha_s m_t^2)$ term can now
be written as a factorized form, where both scales are separated and higher
order QCD corrections of order $\alpha_s^2$ are neglected.

Recalling that pure QCD corrections to $\Gamma(
Z\to b\bar{b})$ are given
by  $(1+\alpha_s(s)/\pi)$ the following
interpretation is at hand.
The virtual top quark is of purely electroweak origin and its effect
can be accounted for by effective vertices.
QCD corrections enter twofold. On the one hand they may be part of the
effective vertex with the relevant scale being $m_t^2$. On the other hand
these QCD corrected vertices are dressed by a virtual gluon, thus
being multiplied by the correction factor $(1+\alpha_s(s)/\pi)$.
The scale of the top mass is in both cases $\mu^2=m_t^2$ because
the overall $m_t^2$ factor always results from the tadpole integral.
We have not found such a
 factorization in the $\ln m_t^2$ term, mainly
because the two classes are not separately finite.

To summrize all contributions discussed in this paper, we use the common
parameters $\rho=1+\delta\rho$ and $\kappa=1+\delta\kappa$ through which
the $Z$ decay rate into two $b$ quarks can be expressed:

\beq\EQN{r5}
\Gamma(Z\rightarrow b\bar{b}) =  \frac{G_F M_Z^3}{12\sqrt{2}\pi}
N_C \rho \left[ 1-\frac{4}{3}\kappa s_W^2 + \frac{8}{9}\kappa^2s^4_W\right]
\eeq

\begin{eqnarray}
\delta\rho^{{\rm\tiny non-univ.}} & = &
 -\frac{G_F}{2\sqrt{2}\pi^2}           \nonumber \\
& &
    \bigg\{
        m_t^2
              \left(1 - \frac{\pi^2}{3}\frac{\as(m_t^2)}{\pi} \right)
              \left(1 + \frac{\as(M_Z^2)}{\pi}\right) \nonumber \\
& &
           +M_Z^2\ln\frac{M_W^2}{m_t^2}
                 \left(-\frac{17}{6}+\frac{8}{3}s_W^2\right) \\
& &
           +\api M_Z^2 \bigg[
                         \ln\frac{M_W^2}{m_t^2}
                             \left( -\frac{17}{6}+\frac{8}{3}s_W^2 \right)
                         +\frac{7}{81} \ln\frac{M_Z^2}{m_t^2}
                            \left(-1+\frac{2}{3}s_W^2 \right)\nonumber\\
& &
\hphantom{+\api M_Z^2}
                         +\Delta\Gamma_{\mbox{\tiny rem}}^{{\tiny OS}}\bigg]
  \bigg\} \nonumber\\
\delta\kappa^{{\rm\tiny  non-univ.}}
                & = & -\frac{1}{2}\delta\rho^{{\rm\tiny non-univ.}}
                                               \nonumber
\end{eqnarray}

\begin{figure}[ht]
 \begin{center}
 \begin{tabular}{c}
   \epsfxsize=12.0cm
   \leavevmode
 \end{tabular}
 \caption{\label{fig2} Leading order and sum of leading and next-to-leading
          order QCD corrections.}
 \end{center}
\end{figure}

The leading $m_t^2$ QCD correction
 is positive in the $OS$ and negative in the
$\msbar$ scheme. In both cases the
corresponding $\ln m_t^2$ term has the
opposite sign.
The leading order $OS$ result is
reduced by about a factor one half. The modulus
of the next-to-leading term in the $\msbar$ scheme is slightly
bigger than the $m_t^2$ term, so that the sum is small and positive.

In Figure \ref{fig2} the leading order and the sum of the leading and
next-to-leading order corrections are plotted against the top mass.
It is easy to see that the inclusion of the next-to-leading
correction reduces the difference of the predictions
between both schemes considerably.

Figure \ref{fig3} compares the QCD corrections with the pure electroweak
result both in the $\msbar$ and in the $OS$ scheme.
One can recognize that in the $\msbar$ scheme the corrections
are numerically tiny ($2.5\cdot 10^{-4}$).

\begin{figure}[ht]
 \begin{center}
 \begin{tabular}{c}
   \epsfxsize=12.0cm
   \leavevmode
 \end{tabular}
 \caption{\label{fig3} Pure electroweak corrections (solid line),
          electroweak + QCD corrections (dashed line).
          The upper curves belong to the $\msbar$ scheme.}
 \end{center}
\end{figure}

To conclude, we calculated QCD corrections to the known electroweak result
for the partial width $\Gamma(Z\to b\bar{b})$ in the limit of a heavy top
quark. We used an expansion in the inverse top mass and calculated
the next-to-leading term of order ${\cal O} (\alpha\alpha_s\ln m_t^2)$.

\vspace{5ex}
{\bf Acknowledgments}

\noindent
We would like to thank K.G. Chetyrkin and J.H. K\"uhn for helpful
discussions.


\end{document}